# Recursive Sketching For Frequency Moments

Vladimir Braverman, Rafail Ostrovsky
University of California, Los Angeles
{vova, rafail}@cs.ucla.edu

October 15, 2018

**Abstract**

In a ground-breaking paper, Indyk and Woodruff (STOC 05) showed how to compute $F_k$ (for $k > 2$) in space complexity $O(\textit{poly-log}(n,m) \cdot n^{1-\frac{2}{k}})$, which is optimal up to (large) poly-logarithmic factors in $n$ and $m$, where $m$ is the length of the stream and $n$ is the upper bound on the number of distinct elements in a stream. The best known lower bound for large moments is $\Omega(\log(n) n^{1-\frac{2}{k}})$. A follow-up work of Bhuvanagiri, Ganguly, Kesh and Saha (SODA 2006) reduced the poly-logarithmic factors of Indyk and Woodruff to $O(\log^2(m) \cdot (\log n + \log m) \cdot n^{1-\frac{2}{k}})$. Further reduction of poly-log factors has been an elusive goal since 2006, when Indyk and Woodruff method seemed to hit a natural "barrier." Using our simple recursive sketch, we provide a different yet simple approach to obtain a $O(\log(m) \log(nm) \cdot (\log \log n)^4 \cdot n^{1-\frac{2}{k}})$ algorithm for constant $\epsilon$ (our bound is, in fact, somewhat stronger, where the $(\log \log n)$ term can be replaced by any constant number of $\log$ iterations instead of just two or three, thus approaching $\log^* n$. Our bound also works for non-constant $\epsilon$ (for details see the body of the paper). Further, our algorithm requires only 4-wise independence, in contrast to existing methods that use pseudo-random generators for computing large frequency moments.

# 1 Introduction

The celebrated paper of Alon, Matias and Szegedy [1] defined the following *streaming* model:

**Definition 1.1.** *Let $m, n$ be positive integers. A* stream *$D = D(n, m)$ is a sequence of size $m$ of integers $p_1, \ldots, p_m$, where $p_i \in \{1, \ldots, n\}$. A frequency vector is a vector of dimensionality $n$ with non-negative entries $f_i, i \in [n]$ defined as:*
$$f_i = |\{j : 1 \le j \le m, p_j = i\}|.$$

**Definition 1.2.** *A $k$-th* frequency moment *of $D$ is defined by $F_k(D) = \sum_{i \in [n]} f_i^k$. Also $F_\infty = \max_{i \in [n]} f_i$.*

Alon, Matias and Szegedy [1] initiated the study of approximating frequency moments with sublinear memory. Their surprising and fundamental results imply that for $k \le 2$ it is possible to approximate $F_k$ with polylogarithmic space; and that polynomial space is necessary for $k > 2$. Today, research on frequency moments is one of the central directions for streaming; many important discoveries have been made since [1]. The incomplete list of relevant work includes [18, 15, 2, 10, 3, 12, 13, 14, 16, 17, 25, 23, 24, 28, 30, 4, 9, 20].

For small $k \le 2$, a long line of papers culminated in the recent optimal results:

- $k = 0$: In their award-winning paper, Kane, Nelson and Woodruff [24] gave optimal-space solution.

- $0 < k < 2$: Kane, Nelson, and Woodruff [23], and later Kane, Nelson, Porat and Woodruff [22], gave optimal-space solutions.

- $k = 2$: The famous sketch of Alon, Matias and Szegedy [1] is, in fact, optimal.

For large $k > 2$, after years of tremendous effort by the theory community, with important intermediate results, the state of the art is as follows:

- $k > 2$ **[Lower bounds:]** The lower bound of $\Omega\left(n^{1-\frac{2}{k}}\right)$ on space complexity was shown by Bar-Yossef, Jayram, Kumar and Sivakumar [2], and Chakrabarti, Khot and Sun [10]. Recently, the lower bound of $\Omega\left((\log n) \cdot n^{1-\frac{2}{k}}\right)$ was announced by Jayram and Woodruff (see the last page of [26] Monemizadeh and Woodruff SODA 2010 presentation of [27]).

- $k > 2$ **[Upper bounds:]** Indyk and Woodruff in their ground-breaking paper [19] first presented a two-pass algorithm with space complexity of $O\left(\frac{1}{\epsilon^{12}} \cdot (\log^2 n)(\log^6 m) \cdot n^{1-\frac{2}{k}}\right)$ and then shown how their two-pass algorithm can be converted to one-pass algorithm with additional poly-log multiplicative factors. The method of Indyk and Woodruff [19] was subsequently improved in 2006 by Bhuvanagiri, Ganguly, Kesh and Saha [5] to achieve: $O\left(\frac{k^2}{\epsilon^{2+4/k}} \cdot (\log^2 m) \cdot (\log n + \log m) \cdot n^{1-\frac{2}{k}}\right)$ space complexity with one pass. To the best of our knowledge, this bound is the best know until today.

**Main Technical Challenge:** No progress was made on the problem of large frequency moments since the 2006 work of [5] described above due to the following "barrier": The large frequency moments represent the case of implicit vectors that cannot be sketched, at least directly. That is, no linear computation is known (unlike the case for the small sketches) that would give a good approximation for the entire vector. In fact, every algorithm that achieves $\tilde{O}(n^{1-2/k})$ memory bits boils down to the Indyk and Woodruff approach. Moreover, this is also true for algorithms for other *implicit* objects [6, 21]. Thus, it might be necessary to not only improve the existing bounds, but also to come up with new methods for computing estimates of implicit vectors.

**Our Results:** This is exactly what we do in this paper. We give a new, *recursive* method of computations of implicit vectors that also improves the upper bounds for large frequency moments. We improve the bound of Bhuvanagiri, Ganguly, Kesh and Saha [5] from $O(k^2 \epsilon^{-2-(4/k)} \log^2(m) \log(nm) n^{1-\frac{2}{k}})$ to at least $O(k^2 \epsilon^{-2-(4/k)} (\log \log(n))^4 \log(m) \log(nm) n^{1-\frac{2}{k}})$. In fact, we give an even better bound. For any constant $t$ we achieve:

$$O\left(\frac{k^2}{\epsilon^{2+(4/k)}} g_t(n) \log(m) \log(nm) n^{1-\frac{2}{k}}\right)$$

space complexity, where:

$$g_0(n) = n$$

and

$$g_t(n) = \log(g_{t-1}(n)).$$

For constant $t$ and $\epsilon$, we can further improve our bound to $O\left(\log(n) \log(n \log(m)) \cdot g_t(n) \cdot n^{1-2/k}\right)$. (Thus, this is a nearly quadratic improvement of the possible ratio between upper and lower bounds compared to the recently announced $\Omega(\log(n) n^{1-2/k})$ lower bound of Jayram and Woodruff.)

Our reduction requires only pairwise independence in contrast to the full independence that previous approaches need. Eliminating the need for total randomness is an important challenge for streaming; see, e.g., [23]. We obtain an algorithm that needs only 4-wise independence and thus does not need Nisan's pseudorandom generators [29]. Finally, we note that our proofs are elementary, along the lines of AMS-type proofs.

**An Alternative Perspective of Our Results:** Many fundamental problems in streaming can be seen as computing $L_1$ approximation of implicit vectors. For instance, the frequency moment $F_k$ can be seen as an $L_1$ norm of a vector with entries $f_i^k$. Except for small moments (i.e., $k \leq 2$), no sketching (i.e., linear transformation) algorithms were known in the past. That is, all previous methods for computing $F_k$ for $k > 2$ resorted to non-linear computations, such as medians to boost the probability that heavy hitters will contribute.

We give a recursive sketching algorithm for estimating within $(1 \pm \epsilon)$ the $L_1$ norm of an *implicit* $n$-dimensional vector of non-negative values, where the algorithm is not given such a vector explicitly, but is only allowed access through a "heavy hitters" oracle. Unlike all previous methods, our recursive sketching algorithm is a *linear* transformation (to heavy hitters) and requires $O(\log n)$ calls to a heavy hitters oracle and yields a $(1 \pm \epsilon)$ approximation to $L_1$ with constant probability. We note that our algorithm can be viewed as a random linear transformation on an *implicit* vector to heavy hitters, and thus gives a new dimension reduction method. Note that our dimension reduction does not contradict the impossibility result of Brinkman and Charikar [8], since our dimension reduction method preserves only the norm of the implicit vector and not pairwise distances between vectors. Yet, our method is sufficient for multiple *streaming* applications where we typically care about the norm of a *single* implicit vector. Thus, we believe that our method might be useful beyond approximating large frequency moments. In particular, it can be applied to other functions and implicit objects such as matrices, e.g., in [6, 21, 7].

**Informal Ideas:** Let us describe, very informally, the fundamental approach of Indyk and Woodruff [19]. They split the frequency vector into "layers," where each layer contains all entries with frequencies between, e.g., $\gamma^i$ and $\gamma^{i+1}$ for a carefully chosen $\gamma > 1$. Then they approximate the contribution of each layer by sampling the stream and by finding the heavy elements that contribute to the layer. Their elegant analysis shows that such a procedure ensures a good approximation with high probability.

We also use the connection between frequency moments and heavy hitters discovered by Indyk and Woodruff. However, we do not use the layers method; we employ recursion instead. For streaming applications, recursion can be helpful if it is possible to reduce computations to a *single* instance of a smaller problem. This is the approach that we take. More specifically, we show that, given an algorithm for "heavy hitters," it is possible to reduce such a problem on a vector of size $n$ to a single computation of a *random* vector of size approximately $\frac{1}{2}n$.

This simple observation follows from elementary arguments such as Chebychev or Hoeffding inequality. We then employ this observation recursively and show that $\log(n)$ recursive calls can give an algorithm that already matches the bounds from [5]. Further, it is possible to reduce the number of recursive calls $\log(n)$ to $\log\log(n)$ by applying the same argument, but stopping after $O(\log\log(n))$ steps. At the depth $O(\log\log(n))$ of the recursion, the number of positive frequencies in a corresponding vector is polylogarithmically smaller then $n$, with constant probability. Thus, any algorithm that works in $polylog(n,m)n^{1-2/k}$ space will approximate such a vector "for free." Employing such an algorithm at the bottom of $\log\log(n)$ recursion reduces the $\log(n)$ factor to a $poly(\log\log(n))$ factor. Further, the same idea may be repeated at least constant number of times; this is how we achieve our final bound. That is, we show that approximating the $L_1$ norm of implicit vectors is practically equivalent to finding heavy hitters. Our method is quite general and works for *any* implicit vector. Further, the simplest variant of the argument requires only pairwise independence, giving an algorithm that requires only 4-wise independence, in contrast to existing methods that use pseudorandom generators.

We gave a simple analysis that uses Chebyshev inequality. Better bounds are possible. For instance, assuming total randomness of $H$ we can apply tail bounds such as the Hoeffding bound or Bernstein inequality. For our purposes, even Chebyshev-like bounds are sufficient, thus we present only these bounds here. Also, pairwise independence allows us to simplify algorithms by avoiding pseudorandom generators.

## 1.1 Roadmap

In Section 2 we introduce the basic argument and extend it to a special case, stuitable for streaming applications, case in Section 3. In Section 4 we describe a generic algorithm for recursive computations. In Section 5 we use our method to obtain a better upper bound for the problem of frequency moments.

## 2 Recursive Sketches

In this paper we denote by $|V|$ the $L_1$ norm of $V$, i.e., $|V| = \sum_{j \in [n]} v_j$.

**Definition 2.1.** *Major elements*
*Let $V$ be a vector of dimensionality $n$ with non-negative entries $v_i \geq 0$. Let $0 < \alpha \leq 1$. An element $v_i$ is a $\alpha$-major with respect to $V$ if: $v_i \geq \alpha|V|$. A set $S \subseteq [n]$ is a $\alpha$-core w.r.t. $V$ if $i \in S$ for any $\alpha$-major $v_i$.*

**Lemma 2.2.** *Let $V \in R^{[n]}$ be a fixed vector and let $S$ be an $\alpha$-core w.r.t. $V$. Let $H$ be a random vector with uniform zero-one entries $h_i, i \in [n]$ that are pairwise-independent. Define*

$$X = \sum_{i \in S} v_i + 2 \sum_{i \notin S} h_i v_i.$$

*Then $P(||X - |V|| \geq \epsilon|V|) \leq \frac{\alpha}{\epsilon^2}$.*

*Proof.* Clearly, $E(X) = |V|$. By the properties of variance, by pairwise independence of $h_i$ and by the definition of $\alpha$-core:

$$Var(X) = 4 \sum_{i \notin S} v_i^2 Var(h_i) = \sum_{i \notin S} v_i^2 \leq \alpha|V|^2.$$

Thus, by Chebyshev inequality:

$$P(||X - |V|| \geq \epsilon|V|) \leq \frac{\alpha}{\epsilon^2}.$$

□

**Corollary 2.3.** *Let $V \in R^{[n]}$ be a random vector and let $S$ be an $\alpha$-core w.r.t. $V$. Let $H$ be a random vector independent of $V$ and $S$ with uniform zero-one entries $h_i, i \in [n]$ that are pairwise-independent. Define*

$$X = \sum_{i \in S} v_i + 2 \sum_{i \notin S} h_i v_i.$$

*Then*

$$P(|X - |V|| \geq \epsilon |V|) \leq \frac{\alpha}{\epsilon^2}.$$

*Proof.* For any fixed $V$ and $S$ the main claim is true since $H$ is independent of $V$ and $S$ and by Lemma 2.2. Thus, the corollary follows. □

### 2.0.1 Recursive Computations

Let $\phi$ be a parameter. Let $H_1, \ldots, H_\phi$ be i.i.d. random vectors with zero-one entries that are uniformly distributed and pairwise independent. For two vectors of dimensionality $n$ define $Had(V, U)$ to be their Hadamard product; i.e., $Had(V, U)$ is a vector of dimensionality $n$ with entries $v_i u_i$. Define:

$$V_0 = V, \text{ and } V_j = Had(V_{j-1}, H_j) \text{ for } j = 1, \ldots, \phi.$$

Denote by $v_i^j$ and $h_i^j$ the $i$-th entry of $V_j$ and $H_j$ respectfully. Let $S_0, \ldots, S_\phi$ be a sequence of subsets of $[n]$ such that $S_j$ is an $\alpha$-core of $V_j$. Define the sequence

$$X_j = \sum_{i \in S_j} v_i^j + 2 \sum_{i \notin S_j} h_i^{j+1} v_i^j, \qquad j = 0, \ldots, \phi - 1,$$

and $X_\phi = |V_\phi|$.

**Fact 2.4.**

$$P(\bigcup_{j=0}^{\phi} (|X_j - |V_j|| \geq \epsilon |V_j|)) \leq \frac{(\phi + 1)\alpha}{\epsilon^2}.$$

*Proof.* Consider fixed $j < k$. It follows from the definitions that $H_{j+1}$ is independent of $V_j$ and $S_j$. Applying Corollary 2.3 and the union bound we obtain the proof. □

Consider the following recursive definition:

$$Y_\phi = X_\phi, \quad Y_j = 2Y_{j+1} + \sum_{i \in S_j} (1 - 2h_i^{j+1}) v_i^j.$$

**Lemma 2.5.** *For any $\phi$, $\gamma$, vector $V$ and $\alpha = \Omega(\frac{\gamma^2}{\phi^3})$:*

$$P(|Y_0 - |V|| \geq \gamma |V|) \leq 0.2.$$

*Proof.* Denote $Err_j^1 = |V_j| - X_j$ and $Err_j^2 = |V_j| - Y_j$. We can rewrite

$$X_j = 2|V_{j+1}| + \sum_{i \in S_j} (1 - 2h_i^{j+1}) v_i^j.$$

Thus $X_j - Y_j = 2(|V_{j+1}| - Y_{j+1}) = 2Err_{j+1}^2$ and

$$|Err_j^2| = |Y_j - |V_j|| \leq |X_j - |V_j|| + |X_j - Y_j| = |Err_j^1| + 2|Err_{j+1}^2|.$$

By definition $Err^1_\phi = Err^2_\phi = 0$. Thus we can rewrite:

$$|Err^2_0| \leq |Err^1_0| + 2|Err^2_1| \leq \cdots \leq \sum_{j=0}^{\phi} 2^j |Err^1_j|.$$

Choose $\epsilon = \frac{\gamma}{10(\phi+1)}$; we have by Fact 2.4:

$$P(|Y_0 - |V|| \geq \gamma |V|) = P(|Err^2_0| \geq \gamma |V|) \leq P(\sum_{j=0}^{\phi} 2^j |Err^1_j| \geq \gamma |V|) \leq$$

$$P\left(\left(\sum_{j=0}^{\phi} 2^j |Err^1_j| \geq \gamma |V|\right) \cap \left(\bigcap_{j=0}^{\phi} (|Err^1_j| < \epsilon |V_j|)\right)\right) + P\left(\bigcup_{j=0}^{\phi} (|X_j - |V_j|| \geq \epsilon |V_j|)\right) \leq$$

$$P\left(\sum_{j=0}^{\phi} 2^j |V_j| \geq 10(\phi+1)|V|\right) + \frac{(\phi+1)\alpha}{\epsilon^2}.$$

For $j > 0$ we note that $|V_j|$ is a random variable defined as:

$$|V_j| = \sum_{i \in [n]} v_i \left(\prod_{t=1}^{j} h_i^t\right).$$

Since all $H_j$ are mutually independent, we conclude that

$$E(\sum_{j=0}^{\phi} 2^j |V_j|) = \sum_{j=0}^{\phi} 2^j \left(\sum_{i \in [n]} v_i \left(\prod_{t=1}^{j} E(h_i^t)\right)\right) = \sum_{j=0}^{\phi} 2^j \left(\sum_{i \in [n]} v_i 2^{-j}\right) = (\phi+1)|V|.$$

Thus, and by Markov inequality, we have

$$P(\sum_{j=0}^{\phi} 2^j |V_j| \geq 10(\phi+1)|V|) \leq 0.1.$$

Also, $\frac{(\phi+1)\alpha}{\epsilon^2} \leq 0.1$ for sufficiently large $\alpha = \Omega(\frac{\gamma^2}{\phi^3})$. Thus,

$$P(|Y_0 - |V|| \geq \gamma |V|) \leq 0.2.$$

□

## 3  An Extension: Approximate and Random Cores

There are many ways to extend our basic result. We will explore one direction, when the cores are random and contain approximations of heavy hitters with high probability[1]. We consider vectors from a finite domain $[m]^n$.

---

[1] In this section we limit our discussion to finite sets and discrete distributions. This limitation is artificial but sufficient for our applications; on the other hand it simplifies the presentation.

**Definition 3.1.** *Let $\Omega$ be a finite set of real numbers. Define $Pairs_t$ to be a set of all sets of pairs of the form:*

$$\{(i_1, w_1), \ldots, (i_t, w_t)\}, \quad 1 \leq i_1 < i_2 < \ldots i_t \leq n, i_j \in N, w_j \in \Omega.$$

*Further define*

$$Pairs = \emptyset \cup \left(\bigcup_{t=1}^{n} Pairs_t\right).$$

**Definition 3.2.** *A non-empty set $Q \in Pairs_t$, i.e., $Q = \{(i_1, w_1), \ldots, (i_t, w_t)\}$ for some $t \in [n]$, is $(\alpha, \epsilon)$-cover w.r.t. vector $V \in [M]^n$ if the following is true:*

1. $\forall j \in [t] (1-\epsilon)v_{i_j} \leq w_j \leq (1+\epsilon)v_{i_j}$.

2. $\forall i \in [n]$ *if $v_i$ is $\alpha$-major then $\exists j \in [t]$ such that $i_j = i$.*

**Definition 3.3.** *Let $\mathcal{D}$ be a probability distribution on $Pairs$. Let $V \in [m]^n$ be a fixed vector. We say that $\mathcal{D}$ is $\delta$-good w.r.t. $V$ if for a random element $Q$ of Pairs with distribution D the following is true:*

$$P(Q \text{ is } (\alpha, \epsilon)\text{-cover of } V) \geq 1 - \delta.$$

**Definition 3.4.** *Let $g$ be a mapping from $[M]^n$ to a set of all distributions on $Pairs$. We say that $g$ is $\delta$-good if for any fixed $V \in [M]^n$ the distribution $g(V)$ is $\delta$-good w.r.t. $V$. Intuitively, $g$ represents an output of an algorithm that finds heavy hitters (and their approximations) of input vector $V$ w.p. $1 - \delta$.*

**Definition 3.5.** *For non-empty $Q \in Pairs$ define $Ind(Q)$ to be the set of indexes of $Q$. Formally, for $Q \in Pairs$, denote $Ind(Q) = \{i : \exists j < t$ such that for $j$-th pair $(i_j, w_j)$ of $Q$ it is true that $i_j = i\}$. For $i \in Ind(Q)$ denote by $w_Q(i)$ the corresponding approximation, i.e. if $i = i_j$ then $w_Q(i) = w_j$. (Note that since $i_j < i_{j+1}$ this is a valid definition.) For completeness, denote $w_Q(i) = 0$ for $i \notin Ind(Q)$ and $Ind(\emptyset) = \emptyset$.*

Now we are ready to repeat the arguments from the previous section.

**Corollary 3.6.** *Let $V \in R^{[n]}$ be a random vector. Let $g$ be a $\delta$-good mapping and let $Q$ be a random element of $Pairs$ that is chosen according to a distribution $g(V)$. Let $H$ be a random vector independent of $V$ and $Q$ with uniform zero-one entries $h_i, i \in [n]$ that are pairwise-independent. Define*

$$X' = \sum_{i \in Ind(Q)} v_i + 2 \sum_{i \notin Ind(Q)} h_i v_i.$$

*Then*

$$P(|X' - |V|| \geq \epsilon|V|) \leq \frac{\epsilon}{\alpha^2} + \delta.$$

*Proof.* Consider a fixed vector $V_0$ and an event that $V = V_0$. Conditioned on this event, the distribution $g(V)$ is fixed and $\delta$-good w.r.t. $V_0$. Consider the event that $Q = Q_0$, where $Q_0$ is an $(\alpha, \epsilon)$-cover w.r.t. $V_0$. Conditioned on this event, $Ind(Q)$ is an $\alpha$-cover w.r.t. $V_0$. Since $H$ is independent of $Q$ the claim is true for any such $V_0$ by Lemma 2.2 and by union bound. Thus, the corollary follows. □

### 3.0.2 Recursive Computations

Let $\phi$ be a parameter. Let $H_1, \ldots, H_\phi$ be i.i.d. random vectors with zero-one entries that are uniformly distributed and pairwise independent. Define:

$$V_0 = V, \text{ and } V_j = Had(V_{j-1}, H_j) \text{ for } j = 1, \ldots, \phi.$$

Denote by $v_i^j$ and $h_i^j$ the $i$-th entry of $V_j$ and $H_j$ respectfully. Let $g$ be a $\delta$-good mapping and let $Q_i$ be a random element of Pairs with distribution $g(V_i)$. Define $w_j(i) = w_{Q_j}(i)$. Define the sequence:

$$X_j' = \sum_{i \in Ind(Q_j)} v_i^j + 2 \sum_{i \notin Ind(Q_j)} h_i^{j+1} v_i^j, \quad j = 0, \ldots, \phi - 1,$$

and $X_\phi' = |V_\phi|$. From Corollary 3.6 and by repeating the arguments from Fact 2.4 we obtain

**Fact 3.7.**
$$P(\bigcup_{j=0}^{\phi} (|X_j' - |V_j|| \geq \epsilon |V_j|)) \leq (\phi+1)(\frac{\alpha}{\epsilon^2} + \delta).$$

Consider the following recursive definition. Let $Y_\phi' = Y_\phi'(V_\phi)$ be a random variable that depends on random vector $V_\phi$ and such that for any fixed $V_\phi$:

$$P(|Y_\phi' - |V_\phi|| \geq \epsilon |V_\phi|) \leq \delta.$$

Also, define for $j = 0, \ldots, \phi - 1$:

$$Y_j' = 2Y_{j+1}' + \sum_{i \in Ind(Q_j)} (1 - 2h_i^{j+1}) w_i^j.$$

**Lemma 3.8.** *For any $\phi$, $\gamma$, vector $V$; for $\alpha = \Omega(\frac{\gamma^2}{\phi^3})$ and $\delta = \Omega(\frac{1}{\phi})$:*

$$P(|Y_0' - |V|| \geq \gamma |V|) \leq 0.2.$$

*Proof.* Denote $Err_j^1 = |V_j| - X_j'$, $Err_j^2 = |V_j| - Y_j'$ and $Err_j^3 = \sum_{i \in Ind(Q_j)} |w_j(i) - v_i^j|$. We can rewrite

$$X_j' = 2|V_{j+1}| + \sum_{i \in Ind(Q_j)} (1 - 2h_i^{j+1}) v_i^j.$$

Thus $|X_j' - Y_j'| \leq 2|Err_{j+1}^2| + |Err_j^3|$ and

$$|Err_j^2| = |Y_j' - |V_j|| \leq |X_j' - |V_j|| + |X_j' - Y_j'| \leq |Err_j^1| + |Err_j^3| + 2|Err_{j+1}^2|.$$

Thus we can rewrite:

$$|Err_0^2| \leq |Err_0^1| + |Err_0^3| + 2|Err_1^2| \leq \cdots \leq 2^k Err_\phi^2 + \sum_{j=0}^{\phi} 2^j |Err_j^1| + \sum_{j=0}^{\phi} 2^j |Err_j^3|.$$

Choose $\epsilon = \frac{\gamma}{30(\phi+1)}$ and denote $Z = 2^k Err_\phi^2 + \sum_{j=0}^{\phi} 2^j |Err_j^1| + \sum_{j=0}^{\phi} 2^j |Err_j^3|$. Then

$$P(|Y_0' - |V|| \geq \gamma |V|) = P(|Err_0^2| \geq \gamma |V|) \leq P(Z \geq \gamma |V|) \leq$$

$$P\left((Z \geq \gamma|V|) \cap \left(\bigcap_{j=0}^{\phi}(|Err_j^1| < \epsilon|V_j|)\right) \cap \left(\bigcap_{j=0}^{\phi}(|Err_j^3| < \epsilon|V_j|)\right) \cap (|Err_\phi^2| < \epsilon|V_\phi|)\right) +$$

$$P\left(|Err_\phi^2| \geq \epsilon|V_\phi|\right) + P\left(\bigcup_{j=0}^{\phi}(|Err_j^1| \geq \epsilon|V_j|)\right) + P\left(\bigcup_{j=0}^{\phi}(|Err_j^3| \geq \epsilon|V_j|)\right).$$

Note that by the definition of $Y'_\phi$, we have $P(|Err_\phi^2| \geq \epsilon|V_\phi|) \leq \delta$. Also, by the definition of $Q_j$ and union bound,

$$P(\bigcup_{j=0}^{\phi}(|Err_j^3| \geq \epsilon|V_j|)) \leq (\phi+1)\delta.$$

Thus and by Fact 3.7:

$$P(|Y'_0 - |V|| \geq \gamma|V|) \leq P\left(\sum_{j=0}^{\phi} 2^j |V_j| \geq 10(\phi+1)|V|\right) + (\phi+2)(\frac{\alpha}{\epsilon^2} + 2\delta).$$

The lemma follows by repeating the concluding arguments from Lemma 2.5. □

## 4 A Generic Algorithm

Let $D$ be a stream as in Definition 1.1. For a function $H : [n] \mapsto \{0,1\}$, define $D_H$ to be a sub-stream of $D$ that contains only elements $p \in D$ such that $H(p) = 1$. Let $V = V(D)$ be an implicit vector of dimensionality $n$ defined by a stream, e.g., a frequency moment vector from Definition 1.1. We say that a vector $V$ is *separable* if for any $H$, we have $Had(V(D), H) = V(D_H)$. Let $HH(D, \alpha, \epsilon, \delta)$ be an algorithm that produces $(\alpha, \epsilon)$-cover w.r.t. $V(D)$ w.p. $1 - \delta$, i.e., produces $\delta$-good distribution w.r.t. $V(D)$ for some suitable finite set of Pairs, as defined in Definition 3.1.

---

**Algorithm 4.1.** *Recursive Sum[0]*$(D, \epsilon)$

1. Generate $\phi = O(\log(n))$ pairwise independent zero-one vectors $H_1, \ldots, H_\phi$. Denote $D_j$ to be a stream $D_{H_1 H_2 \ldots H_j}$.

2. Compute, in parallel, random cores $Q_j = HH(D_j, \frac{\phi^3}{\epsilon^2}, \epsilon, \frac{1}{\phi})$

3. If $F_0(V_\phi) > 10^{10}$ then output 0 and stop. Otherwise compute precisely $Y_\phi = |V_\phi|$.

4. For each $j = \phi - 1, \ldots, 0$, compute $Y_j = 2Y_{j+1} - \sum_{i \in Ind(Q_j)}(1 - 2h_i^j)w_{Q_j}(i)$.

5. Output $Y_0$.

---

**Theorem 4.2.** *Algorithm 4.1 computes $(1 \pm \epsilon)$-approximation of $|V|$ and errs w.p. at most $0.3$. The algorithm uses $O(\log(n)\mu(n, \frac{1}{\epsilon^2 \log^3(n)}, \epsilon, \frac{1}{\log(n)}))$ memory bits, where $\mu$ is the space required by the above algorithm $HH$.*

*Proof.* The correctness follows directly from the description of the algorithm and Lemma 3.8 and Markov inequality. The memory bounds follows from the direct computations. □

# 5 Approximating Large Frequency Moments on Streams

We apply the developed above technique to the problem of frequency moments.

**Fact 5.1.** *Let $V$ be a vector of dimensionality $n$ with non-negative entries and let $n_0$ be a number of non-zero entries in $V$. Let $0 < \alpha < 1$ and let $v_i$ be such that $v_i^k \geq \alpha \sum_{j \in [n]} v_j^k$. Then $v_i^2 \geq 0.5 \alpha^{\frac{2}{k}} n_0^{\frac{2}{k}-1} \sum_{j \neq i} v_j^2$.*

*Proof.* If $n_0 = 0$ the fact is trivial. Otherwise, by Hölder's inequality, $\sum_{j \neq i} v_j^2 \leq n_0^{1-\frac{2}{k}} \left( \sum_{j \neq i} v_j^k \right)^{\frac{2}{k}} \leq n_0^{1-\frac{2}{k}} \alpha^{-\frac{2}{k}} v_i^2$. □

The famous Count-Sketch [11] algorithm finds all $\alpha$-heavy elements. In particular, the following is a corollary from [11].

**Theorem 5.2.** *(from [11]) Let $a_t$ be the frequency of the $t$-th most frequent element. There exists an algorithm that w.p. $1 - \delta$ outputs $t$ pairs $(i, f'_i)$ such that $(1 - \epsilon)f_i \leq f'_i \leq (1 + \epsilon)f_i$ and such that all elements with $f_i \geq (1 - \epsilon)a_t$ appear in the list. The algorithm uses $O((t + \frac{\sum_{i \in [n], f_i \leq a_t} f_i^2}{(\epsilon a_t)^2}) \log(m/\delta) \log(m))$ memory bits.*

Combining with Fact 5.1 we obtain

**Corollary 5.3.** *There exists an algorithm that w.p. $1 - \delta$ outputs $O(\alpha^{-1})$ pairs $(i, f'^k_i)$ such that $(1 - \epsilon)f_i^k \leq f'^k_i \leq (1 + \epsilon)f_i^k$ and such that all elements with $f_i^k \geq \alpha \sum_{j \in [n]} f_j^k$ appear in the list. The algorithm uses $O((\alpha^{-1} + \frac{k^2}{\epsilon^2} \alpha^{-2/k} n^{1-2/k}) \log(m/\delta) \log(m))$ memory bits.*

The algorithm from Corollary 5.3 defines a $\delta$-good distribution w.r.t. to the input vector $V(D)$ over some finite set[2] from Definition 3.1. Denote the algorithm from Corollary 5.3 by $CS(D, \alpha, \epsilon, \delta)$. Thus, combining with Algorithm 4.1 if gives an algorithm errs w.p. $\delta$, outputs $(1 \pm \epsilon)$-approximation of $F_k$ and uses $O(\frac{k^2}{\epsilon^{2+4/k}} n_0^{1-2/k} \log(mn) \log(m) \log^{1+6/k}(n) \log(1/\delta))$ memory bits, nearly matching the bound in [5]. Denote this algorithm by $\mathcal{A}_0(D, \epsilon, \delta)$. We can improve the bound further recursively:

---

**Algorithm 5.4.** *Recursive $F_k[1](D, \epsilon)$*

1. *Generate $\phi = O(\log \log(n))$ pairwise independent zero-one vectors $H_1, \ldots, H_\phi$. Denote $D_j$ to be a stream $D_{H_1 H_2 \ldots H_\phi}$.*

2. *Compute, in parallel, $Q_j = CS(D_j, \frac{\epsilon^2}{\phi^3}, \epsilon, \frac{1}{100\phi})$*

3. *Compute $Y_\phi = \mathcal{A}_0(D_\phi, \epsilon, 0.1)$.*

4. *For each $j = \phi - 1, \ldots, 0$, compute $Y_j = 2Y_{j+1} - \sum_{i \in Ind(Q_j)} (1 - 2h_i^j) w_{Q_j}(i)$.*

5. *Output $Y_0$.*

---

There exists a constant $c$ such that for $\phi = c \log \log(n)$, except with a small constant probability, $F_0(D_\phi) \leq \frac{n}{\log^{10}(n)}$. Thus, executing $\mathcal{A}_0$ for $n' = \frac{n}{\log^{10}(n)}$ we obtain an approximation of $F_k(D_\phi)$ using $O(\frac{k^2}{\epsilon^{2+4/k}} n^{1-2/k} \log(mn) \log(m))$ memory bits. Since $\phi = O(\log \log(n))$, the complexity of the new algorithm becomes $O(\frac{k^2}{\epsilon^{2+4/k}} n^{1-2/k} \log(mn) \log(m) (\log \log(n))^4)$. Repeating this argument a constant number of times we arrive at:

---

[2]Indeed, we can define the finite set $\Omega$ from Definition 3.1 as a set of all possible outputs of Count-Sketch executed over all vectors on $[m]^n$. This is a finite set (for finite $n, m$) and thus we can define Pairs accordingly.

**Theorem 5.5.** *Define $g_1(n) = \log(n)$ and $g_t(n) = \log(g_{t-1}(n))$. For any constant $t$ there exist an algorithm computes a $(1 \pm \epsilon)$-approximation of $F_k(D)$, errs w.p. at most $\frac{1}{3}$ and uses $O(c_t k^2 \epsilon^{-2-(4/k)} n^{1-\frac{2}{k}} g_t(n) \log(m) \log(nm))$ memory bits, where $c_t$ is a constant that depends on $t$.*

We note also that it is possible to reduce the complexity to $O(n^{1-\frac{2}{k}} g_t(n) \log(n)(\log(n) + \log\log(m)))$, at least for constant $\epsilon$, using, instead of CountSketch, the variant of the AMS sketch and the ideas from [7].